%% file: sigori.tex
\documentclass[usenatbib]{mn2e}
\usepackage{graphicx}
\usepackage{lscape}
\author[Ben Burningham et al.]{Ben Burningham$^1$, Tim Naylor$^1$,
  S. P. Littlefair$^1$, R. D. Jeffries$^2$ \\
$^1$ School of Physics, University of Exeter, Stocker Road, Exeter EX4 4QL\\
$^2$ Department of Physics, Keele University, Keele, Staffordshire ST5 5BG}

\title[Contamination and exclusion in $\sigma$ Orionis]{Contamination
  and exclusion in the $\sigma$ Orionis young group.}

\begin{document}
\include{aas_macros}

\linespread{1.3}
\maketitle

\begin{abstract}

We present radial velocities for 38 low-mass candidate members of the
$\sigma$ Orionis young group.  We have measured their radial velocities by
cross-correlation of high resolution (R$\approx$ 6000) AF2/WYFFOS
spectra of the gravity sensitive Na {\sc i} doublet at 8183, 8195~\AA.  
The total sample contained 117 objects of which 54 have sufficient
signal-to-noise to detect Na {\sc i} at an equivalent width of 3~\AA,
however we only detect Na {\sc i} in 38 of these.  
This implies that very low-mass members of this young group display weaker
Na{\sc i} absorption than similarly aged objects in the Upper Scorpius OB
association.  
We develop a technique to assess membership using
radial velocities with a range of uncertainties that does not bias the
selection when large uncertainties are present.  The resulting
membership probabilities are used to assess the issue of exclusion in
photometric selections,  and we find that very few members are likely to
be excluded by such techniques.
 We also assess the level of contamination in the
expected pre-main sequence region of colour-magnitude space brighter
than I = 17.  
We find that contamination by non-members in the expected PMS region of the
colour-magnitude diagram is small.  We
conclude that although radial velocity alone is insufficient to
confirm membership, high signal-to-noise observations of the Na{\sc i}
doublet provide the opportunity to use the strength of Na{\sc i} absorption in
concert with radial velocities to asses membership down to the lowest
masses, where Lithium absorption no longer distinguishes youth.

\end{abstract}


\section{Introduction}
Studies of the low-mass and sub-stellar initial mass function (IMF) provide the
opportunity to constrain theories of star formation and investigate
environmental influences on the final results of the formation
process. 
Whilst the high and intermediate mass regions of the IMF have
been well studied and found to be essentially universal in the field
and in clusters
\citep[e.g.][and references therein]{h2003,gcc82}, this is not the
case for low and sub-stellar masses. 
Although there have been a number of studies of the low-mass
IMF in the field \citep[e.g.][]{rklbgbdmcbs99, k2001}, due to their inherent
faintness only the most nearby of field T dwarfs are detected, leading
to poor statistics at these masses.  
Searches for brown dwarfs in intermediate age ($10^8$ yrs) clusters
such as the Pleiades have proved fruitful \citep[e.g.][and references
  therein]{mbsc2003, b2000}. 
Unfortunately, to derive an IMF from the mass function of an
intermediate age cluster requires corrections for the effects of
stellar and dynamical evolution over the age of the cluster.

Young clusters and  associations are the ideal places to study this region
of the IMF.  This is because many of the problems associated with
measuring the IMF at the lowest masses disappear in such young
regions.  Firstly no correction need be made for the effects of
stellar or dynamical evolution, the regions are sufficiently young that the
measured mass function {\it is} the IMF. Also, the objects at
substellar masses are much easier to observe at ages of less than 10 Myrs
since they still shine brightly as gravitational potential is
released.  In addition, a supposed advantage of clusters and associations
with ages of less than 2 Myrs is the lack of
contaminating sources due to the fact that such young clusters are
often still associated with their natal molecular cloud, which will
tend to obscure the background to high extinctions.
The Orion Nebula Cluster (ONC) has been extensively studied for just
these reasons \citep[e.g.][]{mlla2002,lryccrst2000, lr2000}. 
However, inspection of the methods which must be employed for such
studies reveals that association with a molecular cloud brings with it
its own, unavoidable, problems.

Since there is often a spread of ages in young star forming regions,
it is preferable to use photometric magnitudes and spectroscopically
derived temperatures in conjunction with pre-main sequence (PMS)
models to determine the age and mass for each object in the sample.
Unfortunately due to the high extinctions present along sightlines
toward young, partially embedded, clusters it is generally not
possible to obtain spectroscopy for the lowest mass objects.  Since
there is also normally a large range of reddening toward objects
within such clusters it is not possible to determine the age and mass
for each object, and a distribution of ages and reddening must be
assumed to obtain the IMF from the luminosity function (LF)
\citep{mlla2002}.  
Another effect of the high extinction is that LFs
must be constructed in the K- and L-band to avoid biasing the sample
to the least extincted objects and to
ensure a good level of completeness. However, it has been observed in
the young clusters such as the ONC that up to 85\% of the
objects have circumstellar discs, which are responsible for infra-red excesses
 \citep[e.g.][]{lmhlatw2000}, making mass functions
derived from K- and L-band LFs unreliable.  
These problems are avoided if star forming regions between 2 and 10
Myrs old are used for IMF studies.  At these ages the clusters have often
emerged fully from their natal cloud, presumably due to the UV flux
from the massive stars evaporating the molecular gas.  
The lower extinction allows the
luminosity function to be constructed in the I-band, which is
less affected by circumstellar discs than the K-band, in addition to making
spectroscopy possible down to very low masses.
Since we need such regions to be
as nearby as possible, to allow studies of the lowest mass objects,
the number of prospective laboratories is very small.  The most
promising candidate is the $\sigma$ Orionis ($\sigma$ Ori) young
group, a constituent of the Orion OB1b association, which has age
estimates in the range 1.7 - 5 Myrs \citep{wh78,bgz94}. 
The low extinction, $E(B-V) = 0.05$ \citep{lee68}, and small distance,
$D\approx  350$pc \citep{hipp97}, make it ideal for studies of
the low mass IMF.  

\citet{bejar2001} have carried out an investigation of the substellar IMF
in the $\sigma$ Ori region.  They used a sample of 9
spectroscopically confirmed members to define the locus of the cluster
sequence between M6 and L4 spectral types.  They then used the location
of this sequence in colour-magnitude (C-M) space to identify cluster
members, and construct an I-band LF.  
Using several different models to construct their IMF,
\citet{bejar2001} found a mass spectrum $(dN/dm) \propto m^{-\alpha}$
with an exponent of $\alpha=$0.8$\pm$0.4 between 0.2 and 0.01M$_{\odot}$, 
i.e. the number of objects per mass bin continues to
increase with decreasing mass down to below the deuterium burning
limit and into the realm of planetary mass objects.  This corresponds
to an IMF $\xi(m) \propto M^{-\gamma}$ with an exponent $\gamma$ =
-0.2.
The low extinction, however, means that background contamination
cannot be ruled out, even in the locus of the cluster sequence.
\citet{barrado2001} obtained spectral types for 
roughly one quarter of the photometrically selected candidates,
confirming a high proportion of them as cool objects.  Whilst such a
method is effective at removing reddened background objects from a
sample, it is not robust against interloping field dwarfs.  

When the
number of objects in each mass bin is so small, however, a small
number of contaminants can have a significant effect on the derived
mass function.  
Whilst \citet{bejar2001} are careful with their
photometric selection to avoid contamination, this could, in itself,
lead to poorly understood completeness.  
Without empirically determining how far into the expected background
region of the colour-magnitude diagram (CMD) bona fide members are
still found, any photometric selection which minimises contamination
may well exclude a significant fraction of member objects.

The likelihood of age spread
in such a young cluster makes this consideration crucial.
\citet{kenyon2004} obtained fibre spectroscopy for a sample
of over 70
candidate members, drawn from a $RI$ catalogue, with masses between
0.055 and 0.3M$_{\odot}$ (they assume a cluster age of 5 Myrs). They
found that contamination from background objects was limited for the
reddest region of each magnitude bin in the I/R-I CMD, but the PMS
objects were poorly correlated with isochrones in this colour.  They
found better correlation in I/I-J, the colours used by
\citet{bejar2001},  but the contamination was found to be considerably
worse, becoming more prevalent at the faintest magnitudes.  
Additionally, the CMDs of \citet{kenyon2004} show that any
strict cut along isochrones  in either set of colours would exclude a
significant number of members.  

It is clear that the only way to obtain a reliable IMF for this region
is to use a spectroscopically confirmed sample of members, drawn from
a broad region of colour magnitude space.
\citet{kenyon2004} used the
presence of Li I absorption at 6708~\AA\ to confirm youth and thus
membership for objects in their sample.   This technique, however, is
limited to objects with M $\geq$ 0.065M$_\odot$. Since objects below
this mass never attain high enough core temperatures to burn Lithium,
old foreground T-dwarfs cannot be distinguished from bona fide members
based on Li absorption alone. 
 
If a reliable IMF is to be derived for the lowest masses, another
method of distinguishing members must be found.  It is with this in
mind that we have obtained fibre spectroscopy of the Na{\sc i} doublet
at 8183, 8195~\AA\  for a photometrically selected
sample of objects in the direction of the $\sigma$ Ori cluster.  Using
this doublet has several benefits.  Cool objects such as low-mass
stars and brown dwarfs are brightest in the NIR and IR regions, where
night sky emission can severely affect observations.  The Na{\sc i} doublet
lies at a wavelength which is relatively unaffected by bright sky
emission lines.  Also the Na{\sc i} doublet is a relatively strong spectral
feature in cool stars, and so should be easy to observe a moderate
signal-to-noise.  

In this paper we have two aims. Firstly, to investigate the use of radial
velocity, as measured from the Na{\sc i} doublet, as a membership
diagnostic. Whilst Li absorption would be an effective diagnostic for
much of our sample, this work is concerned with developing a
spectroscopic technique for use at magnitudes and colours where
foreground contaminants may lie below the Li burning limit. 
Our second aim is to measure the level of contamination from non-members in the
expected PMS region of the CMD and to determine if many bona fide
members are excluded by photometric selection techniques.

The paper
will be laid out as follows.  In Section~\ref{sec:specs} we will discuss the
rationale behind our target selection and our observations.  In
Section~\ref{sec:data}
 we will describe our data reduction technique and our method
for sky subtraction.  In Section~\ref{sec:members} we will describe
our sample selection, the cross
correlation of spectra to obtain radial velocities and the calculation
of membership probabilities.   
Section~\ref{sec:disc} will contain a discussion of the results.  A
summary of our conclusions will be given in Section~\ref{sec:conc}.

\section{Optical Spectroscopy Using AF2/WYFFOS}
\label{sec:specs}
\subsection{Target Selection}
\label{sec:targets}

\begin{figure}
\includegraphics[height=375pt,width=275pt, angle=90]{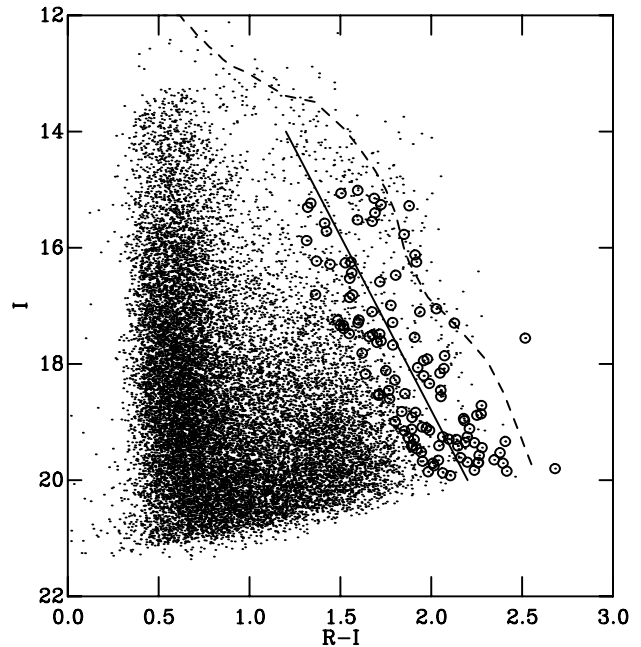}
\caption{The CMD for the optical $RI$ catalogue from which targets
  were selected \citep{kenyon2004}.  Targets for this survey are shown
  as circles.  The dotted line follows a NextGen 5 Myr isochrone
  \citep{cb97, bcah2002}.  We define the expected background region to
  be blueward of the solid line, and the PMS region to be redward of
  it (see section 5).}
\label{fig:sample}
\end{figure}

\begin{figure}
\includegraphics[height=325pt,width=275pt, angle=90]{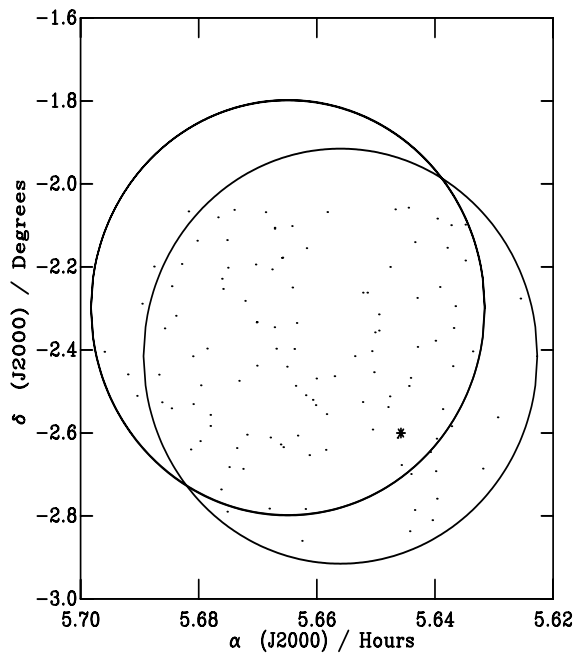}
\caption{The locations of our targets and WHT FoVs.  $\sigma$
  Ori is shown as star, whilst our targets are shown as circles.  The
  WHT FoVs are shown as large circles. }
\label{fig:fov}
\end{figure}

We obtained spectra for 117 objects using the AF2/WYFFOS fibre
spectrograph on the William Herschel Telescope at the Observatorio del
Roque de Los
Muchachos, La Palma.  The targets were selected from the photometric
$RI$ catalogue described by \citet{kenyon2004} and were chosen to
compliment the sample for which spectroscopic observations have
already been obtained \citep{kenyon2004}. 
Since many of the reddest objects have
already been observed, particularly at the brighter magnitudes, we
were able to cut deep into the region which is thought to be background
contamination, see Figure~\ref{fig:sample}.  This
method of target selection has served a dual purpose: 1) it allows us
to investigate the possibility of contamination in the
PMS region of colour magnitude space; 2) it allows us to investigate
the number of member objects which might be lost in more
conservatively drawn samples. 
The distribution of the targets, in relation to $\sigma$ Ori, and the
WYFFOS/WHT FoV is shown in Figure~\ref{fig:fov}.

\subsection{Observations}
\label{sec:obs}

We observed our targets using the small fibre
(1.6'') setup on the nights of the 3$^{rd}$ and 4$^{th}$ January 2003.
Conditions were dark, and cloud was only present during the first half
of the second night.  We observed in 3rd order echelle mode, with a
central wavelength of 8300~\AA.  This gives a dispersion of 0.57
~\AA/pixel and we achieved a resolution of approximately 1.4~\AA.  
We elected to observe the Na{\sc i} doublet at high resolution so as to
avoid broadening the bright sky emission lines and obscuring the doublet.

 We achieved a total exposure time of 9.5 hours for the faintest
 targets. This time was split between 4 configurations to maximise the
 number of objects we could observe (see Table~\ref{table:exp}). 
The brightest targets were only observed in the shortest set-ups, with
 fainter ones observed using a combination of set-ups to achieve 
 appropriate integration times.  Each set of exposures was
subdivided into Sections of 900s or 1800s, depending on the total
exposure required for the set-up. Between exposures of the target
objects, the telescope was moved such that $\sigma$ Ori lay over a
fibre, and a spectrum of $\sigma$ Ori was obtained for the purposes of
telluric correction.  Copper-Neon arc frames were also obtained for
 both positions of the telescope at these times.
To maximise our exposure times 
we did not observed offset sky fields, as is advised by \citet{wg92}, 
 for estimating the relative throughput of the fibres prior to sky
 subtraction.
However, the relative throughput of the fibres in the small fibre bundle is
considerably more homogeneous than it had been for the older large
fibre bundle (see
http://www.ing.iac.es/Astronomy/instruments/af2/index.html), and we
have an alternative method for achieving sky subtraction, which will
be described later.  We also observed several stars for use as radial
 velocity standards, these observations are detailed in
 Table~\ref{table:standards}, along with the identifier of the
 telluric reference observed.

Unfortunately, many of the spectra we obtained suffered from poor
signal-to-noise, and others had no detectable signal at all. 
We have correlated the $RI$ photometric catalogue used to select our targets 
with a 2MASS catalogue for the same region and find the RMS in
the residuals of the coordinates to be approximately 0.2'' in both
axes.  As such, we rule out errors in the astrometry of the $RI$
catalogue as an explanation for our missed targets.   
We also rule out proper motion as a factor.  At the target selection
stage the 2MASS catalogue for our potential targets was correlated
against the SuperCosmos sky survey catalogue for the same region and
objects with large coordinate residuals were excluded.
Whilst source variability, well known in T Tauri stars, may
have had an influence on the signal detected from some of the targets,
we believe the failure to achieve the desired signal-to-noise in so
many spectra is the result of two problems with the AF2/WYFFOS field
of view (FoV).
Firstly, the point spread function (PSF) is highly variable across the FoV,
becoming very smeared out beyond a radius of 25 arcminutes.  
As such,
when using the small fibre bundle a large proportion
of the signal from a target star in the outer area of the FoV will be
lost.  This effect explains the poor signal-to-noise found in the
spectra of some objects. 
The second problem is that the astrometric distortion in the
AF2/WYFFOS FoV is poorly constrained, meaning that many objects were
simply missed by their assigned fibre.
The method employed for selecting our sample
from the available spectra is discussed in Section~\ref{sec:sample}.

\begin{table}
\caption{Summary of the exposure times and sky fibre allocations for
  the 4 configurations used.}
\label{table:exp}
\begin{tabular}{c c c c} 
\hline
 Configuration & Integration Time (s) & No. of Sky Fibres \\
\hline \hline

Long1 & 12600 & 58 \\
Long2 & 12600 & 49 \\
Medium & 5400 & 35 \\
Short & 3600 & 29 \\
\hline
\end{tabular}
\end{table}

\section{Data Reduction}
\label{sec:data}

\subsection{Extraction of Spectra}
\label{sec:extract}
The spectra were extracted and dispersion corrected using the WYFFOS
data reduction routines within the RGO package within the IRAF
environment \citep{lewis96}.  The spectra were bias subtracted using
the mean bias level in the overscan region of the Tek6
detector. Flatfielding was achieved using Tungsten flatfields, which
were also used for aperture detection.  The apertures were traced using
steps of 5 pixels and fitted using a 2$^{rd}$ order spline function.
Spectra were then optimally extracted from target, $\sigma$ Ori and
arc frames.  
Dispersion correction  for each spectrum was
achieved by first fitting a 4$^{th}$ order Chebyshev function to arc
lines that were manually identified in the spectrum from the middle
aperture of the appropriate arc frame.   
An automated procedure then obtained solutions using the same lines in the
remaining apertures.  The RMS deviation from the dispersion solution
across the apertures in the arc frames was typically 0.01-0.02~\AA. 
 These solutions were then applied to spectra
from the same apertures in the target frame.

\subsection{Telluric Correction and Sky Subtraction}
\label{sec:sky}
Absorption features due the Earth's atmosphere can influence the
results of cross-correlation, by biasing the result towards zero
velocity.  As such the effects of the atmospheric, or telluric,
absorption must be corrected for.  It is normal for an O-star, whose
spectrum should be nearly featureless, to be used to as a
reference spectrum for telluric correction.  This can lead to
difficulty when the reference star is well separated from the targets,
as one is required to assume that the telluric absorption is uniform in
both time and position.  
Fortunately, $\sigma$ Ori itself provides us with an appropriate reference
spectrum, and is located within our FoVs.  As such, we need only
assume that the telluric absorption is uniform over short periods of time.

Telluric correction was carried out on spectra from each sub-exposure
using the temporally closest $\sigma$ Ori spectrum.  Radial
velocity standard spectra were corrected using a telluric reference observed
immediately before or after each exposure, and chosen to be close-by
on the sky, and these are listed in Table~\ref{table:standards}.  
In each case the object spectrum was divided by a
rectified version of the telluric reference spectrum, thus removing
absorption features originating in the Earth's atmosphere.

The object spectra from the sub-exposures of each configuration were
combined by median stacking prior to sky subtraction.  We calculated
the uncertainties in flux at this point as the standard error about
the median.
Two master sky spectra were constructed for
each configuration by calculating (i) the weighted mean and (ii) the median of
median stacked spectra from each sub-exposure.  The sky spectra used
for constructing the median sky for each sub-exposure were selected on
the basis of being free of cosmic rays in the region of interest
(8100~\AA\ $\leq \lambda \leq$ 8300~\AA). Since our target region is
well seperated from the nebular emission near the Horsehead, the
background is flat.  As such, use of a master sky spectrum for each
set-up is appropriate (see below also).

\begin{figure} 
\includegraphics[height=250pt,width=190pt, angle=90]{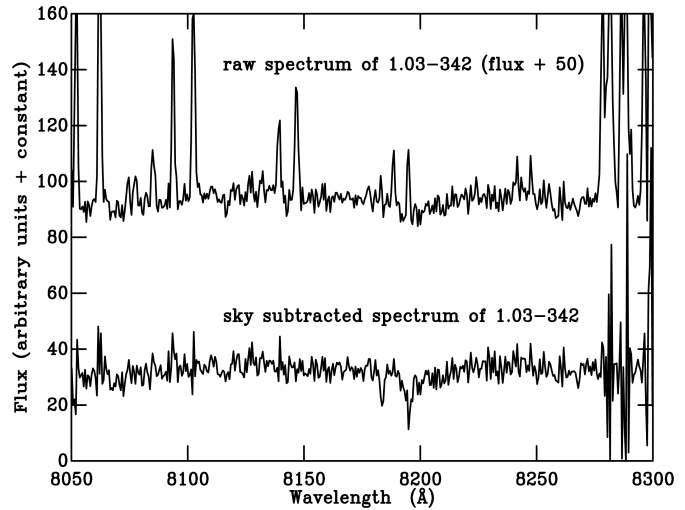}
\caption{An example of the effectiveness of our method for sky
  subtraction.  In this case, the median stacked exposures from 3.5
  hours of the 7 hour total exposure time for our target 1.03-342 are
  sky subtracted using the method described above.}
\label{fig:ssub}
\end{figure}

\begin{figure} 
\includegraphics[height=250pt,width=190pt, angle=90]{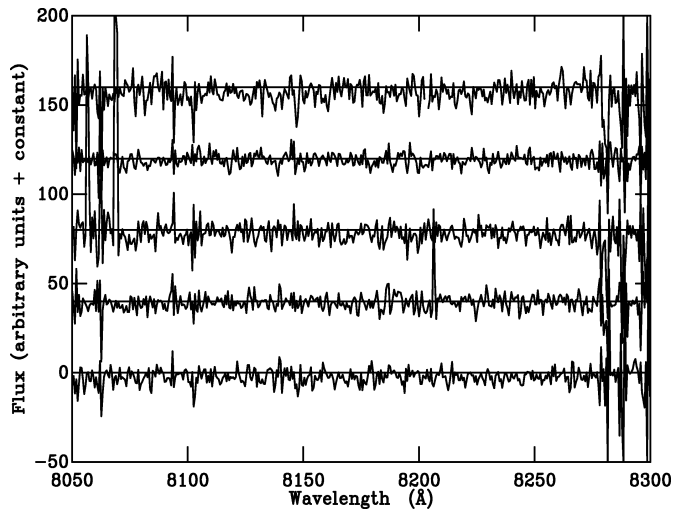}
\caption{Examples of sky subtracted sky spectra used to check validity
of using the diagnostic lines near 8150 \AA }
\label{fig:skies}
\end{figure}

  Since no offset sky frames were obtained, it
was not possible to estimate the relative throughput of each fibre
and use this to scale the master sky spectrum to each fibre for
subtraction.  
Instead we have iteratively scaled the master sky frame until the
subtraction is optimised for the removal of sky lines.  The method we
use is identical to that used by \citet{bsnh2000} for removing the,
relatively sharp-spiked, secondary star spectrum from that of a
cataclysmic variable (CV). 
In this case the master sky frame is treated as a template secondary
star spectrum, and the target+sky as the CV spectrum.
First we subtract the master sky frame with no scaling and smooth the
residual with a smoothing length of 50 pixels.  The smoothed residual
is then subtracted from the unsmoothed residual and the $\chi^{2}$ per
pixel calculated.  This procedure is then repeated with a  series of different
scale factors.  The scale factor that results in the lowest value of
$\chi^{2}$ is the one which is then applied to master sky spectrum
prior to subtraction from the data. An example of the success of this
method of sky subtraction is shown in Figure~\ref{fig:ssub}.  
This method was carried out using both versions of the master sky
frames.  As can be seen in Figure~\ref{fig:ssub},
sky emission lines lie at similar wavelengths to the Na{\sc i} doublet of
interest. As such it is crucial that the sky lines are not
over-subtracted, giving rise to spurious absorption features.  To
assess the effectiveness of each sky subtraction we used the two
emission lines just blueward of 8150~\AA\ as diagnostic lines.  Since
these lines were found to be more intense than those situated over
Na{\sc i} doublet in the spectra from all sky fibres,
they are good indicators of over-subtraction. A problem would arise if
the relative strengths of these lines varied significantly across the
field of view, or over the period of time that a set-up was observed
for. To assess this potential problem we have tested the sky
subtraction on a number of sky fibres from across the field of view
and from different exposures in a set-up.  No cases were found where
over-subtraction had taken place but was not evidenced by absorption
in the diagnostic lines.  Figure ~\ref{fig:skies} shows a selection of
these test subtractions to illustrate this point.

The visual
inspection of the residual spectra resulted in a preferred version of
the sky frames to be used for each object in a configuration, and this
was forced to be the same across all configurations for a given object
to ensure a consistent method.  
In the vast majority of cases the median stacked sky frames performed better.
It should be noted that since the spectra have not been corrected for the
relative throughput of the fibres, it is not possible to flux
calibrate spectra that have been sky subtracted in this manner. 
At this
stage spectra for objects obtained in different configurations were
co-added to arrive at the final spectra.  These spectra are available
on-line via the CDS service or from the Cluster collaboration
homepage\footnote{http://www.astro.ex.ac.uk/people/timn/Catalogues/description.html}.

\section{Selecting Members by Radial Velocity}
\label{sec:members}

\subsection{Sample Selection}
\label{sec:sample}

As discussed in Section~\ref{sec:targets}, we did not obtain good
data for every object in the sample.  To avoid biasing our sample at
this stage we opted to apply a signal-to-noise cut to select our
sample. We measured the equivalent width of the Na{\sc i}
doublet, hereafter EW(Na{\sc i}), in all the spectra obtained.  
We used two continuum
bands: 1) 8151-8173~\AA; 2) 8233-8255~\AA.  The wavelength range for
continuum band 2 was selected to avoid the TiO band which lies redward
of 8200~\AA.  We integrated EW(Na{\sc i}) over the range 8180-8202~\AA.    
Those objects which had a sufficiently
small uncertainty to allow a 2$\sigma$ detection of the doublet at an
equivalent width (EW) of 3~\AA, i.e. $\sigma \geq$ 1.5~\AA, were included in
the sample.  This EW was chosen as most brown dwarfs
detected by \citet{mdg2004} in the Upper Scorpius OB association
displayed an EW(Na{\sc i}) above this value.  
 The selected objects and their values for EW(Na{\sc i})
are listed in Table~\ref{table:results}.
Our objective measure for the presence of Na{\sc i} absorption was an
EW(Na{\sc i}) measured at a significance of 2$\sigma$ or greater.  Our
objective measure failed to detect Na{\sc i} in 11 of the
selected sample objects.  We rule out the presence of Na{\sc i} in a
further 5 objects, despite the objective measure suggesting its
presence.  These false detections were caused in two cases (1.03-460,
1.03-612) by the small sky line residuals affecting the continuum
estimate.  
One false detection (8.03-396) was caused by over-subtracted sky lines
within the EW integration band.  
The other two false detections (8.04-50, 8.04-77) were
caused by unidentified absorption features encroaching on the EW
integration band.
Since no radial velocity can be obtained
when Na{\sc i} is not detected, we do not include such objects in our cross
correlation nor do we calculate membership probabilities.
 
All of the objects with a failed Na{\sc i} detection are fainter than I=17.  
The strength of the Na{\sc i}
doublet is gravity dependant, getting stronger with increasing surface
gravity. As such it has been used as a discriminator between field dwarfs and
giants \citep[e.g.][]{sbrm97} and field dwarfs and young brown dwarfs
\citep{mdg2004}. 
Failure to detect Na{\sc i} in the spectrum of an object could
indicate that it is a giant with a sufficiently low surface gravity to
weaken Na{\sc i} beyond our detection threshold.
However, since very young brown dwarfs
also have low surface gravity the same might be expected from members of the
$\sigma$ Ori group.  Surface gravities are typically higher in young
(age $\leq$ 5 Myrs) 
brown dwarfs, $log(g) \sim 3.5$, than in M-giants,
$log(g) \sim 0$ \citep{gmrl2003}, so we might expect to be able to
distinguish between extremely low surface gravity giants and young
brown dwarfs.  This will be discussed further in Section~\ref{sec:disc}.

\subsection{Cross Correlation of Spectra to Obtain Radial Velocities}
\label{sec:xcor}

We obtained radial velocities for objects in our sample which
displayed Na{\sc i} absorption  by cross
correlating their spectra against those of stars with known velocities
and similar spectral types to our targets.  
The stars observed as radial velocity standards are detailed in
table~\ref{table:standards}.  
The wavelength range used for the cross correlation was restricted to
8175 - 8205~\AA\ to minimise the effects of noise in the spectra.
We restricted the range of acceptable correlations to be
$\pm$100km s$^{-1}$, and as a result velocities were not found for some
objects on some cross correlations.  Uncertainties for each velocity
measurement were estimated by perturbing the value of the flux at each data
point in the target spectrum by some value drawn from a Gaussian based
on the uncertainty in the flux.  This was repeated 100 times and the
resulting spectra were cross correlated against the standards in the
same manner as the original data.  The uncertainty in velocity was taken to be
the standard deviation in the distribution of velocities from the
perturbed spectra. The quality of all cross correlation functions
were assessed by visual inspection and none were found to have any
evidence of the multiple peaks that might be expected if the sky
subtraction introduced spurious features.

Although the 4 different velocity standards gave similar
results, the use of GJ3517 gave the most successful correlations.  As
such, this is the one we chosen to use for further analysis.
The velocities and uncertainties resulting from cross correlations against GJ
3517 are given in Table~\ref{table:results}.

\begin{table*}
\caption{Details of stars observed as radial velocity standards and
  details of their observations. Radial velocities listed are
  barycentric. Equivalent widths for Na{\sc i} are
  measured with continuum bands at (8151-8173~\AA) and
  (8233-8255~\AA), and integrated over the range (8180-8202~\AA).}
\label{table:standards}
\begin{tabular}{c c c c c c c c c c} 
\hline
Star & $\alpha$ (J2000) & $\delta$(J2000) & Sp. Type & I &
Integration time (s)& $V_{rad}$ / km s$^{-1}$  & EW(Na{\sc i})
/ \AA\ & $\sigma_{EW}$ &  Telluric reference \\
\hline \hline
WX UMa & 11 05 30.31 & +43 31 16.6 & M6 &  & 200 & 68$^1$ & 8.28 & 0.10 & SAO 43460\\
GJ412A & 11 05 28.58 & +43 31 36.4 & M0.5 &  & 10 & 68$^1$ & 2.18 &
0.09& SAO 43460\\
GJ3517 & 08 53 36.11 & -03 29 32.4 & M9 & 14.44 & 400 & 9$^2$ & 5.04 & 0.38 & SAO 136434 \\
LP213-67 & 10 47 12.65 & +40 26 43.7 & M6.5 & 13.17 & 600 & 5$^3$ & 6.55 & 0.21 & SAO 62257 \\
\hline
\end{tabular}
References for radial velocities: 1) \citet{dfpm98}; 2) \citet{tr98};
3)\citet{rklgdm2002}.
\end{table*}

\begin{table*}
\caption{Equivalent widths and radial velocities measured using the Na{\sc i} doublet.  The field number
  and object number refer to designations found in
  \citet{kenyon2004}. Quality flags for I and R-I are as described in
\citet{me2003}. The continuum was measured from bands at 8151-8173~\AA\
  and 8233-8255~\AA.  The EWs were integrated over the range
  8180-8202~\AA. 
Radial velocities were measured by cross correlating against
  GJ3517, and corrected to give barycentric radial velocities. }
\label{table:results}
\begin{tabular}{c c c c c c c c c c c c c c c c c c c c} 
\hline
Field & Object & $\alpha$(J2000) & $\delta$(J2000) & $I$ &
$\sigma_I$ & Flag & $R-I$ & $\sigma_{R-I}$ & Flag & EW &
$\sigma_{EW}$ & $V_{bar}$ & $\sigma_{V}$  & P$_{vel}$ \\
&&&&&&&&&& \AA & \AA & km s$^{-1}$ & km s$^{-1}$ & \% \\
\hline \hline

  1.01  &   319   &  05 39 48.911 & -02 29 11.05 &    15.012 &      0.007 &  VV &    1.595 &      0.011 &  VV &    3.23 &  0.11 & 24.25 & 0.86 & 77.3 \\
  8.02  &   179   &  05 39 47.696 & -02 36 22.96 &    15.064 &      0.005 &  OO &    1.508 &      0.008 &  OO &    3.04 &  0.10 & 26.37 & 1.13 & 99.9 \\  
  4.03  &   229   &  05 38 23.536 & -02 41 31.66 &    15.149 &      0.008 &  OO &    1.686 &      0.013 &  OO &    2.31 &  0.25 & 27.61 & 2.38 & 98.8 \\
  1.01  &   253   &  05 40 51.369 & -02 31 49.93 &    15.232 &      0.007 &  OO &    1.337 &      0.011 &  OO &    3.21 &  0.30 & 10.48 & 4.91 & 0.0 \\
  8.02  &   143   &  05 39 56.445 & -02 38  3.43 &    15.258 &      0.005 &  OO &    1.725 &      0.007 &  OO &    2.17 &  0.11 & 31.18 & 1.00 & 100.0 \\
  3.01  &    67   &  05 38 46.835 & -02 36 43.38 &    15.278 &      0.016 &  OO &    1.878 &      0.038 &  OO &    0.64 &  0.10 & 32.11 & 1.32 & 100.0 \\
  2.03  &   260   &  05 39 30.561 & -02 38 26.89 &    15.304 &      0.007 &  OO &    1.320 &      0.011 &  OO &    3.84 &  0.12 & 18.98 & 1.12 & 0.00 \\
  4.03  &   237   &  05 37 54.857 & -02 41  9.15 &    15.400 &      0.006 &  VV &    1.690 &      0.010 &  VV &    1.32 &  0.39 & 33.86 & 3.56 & 89.5 \\
  1.03  &    60   &  05 40 30.179 & -02 12  6.13 &    15.520 &      0.007 &  VV &    1.594 &      0.011 &  VV &    4.13 &  0.76 & 31.87 & 6.26 & 84.3 \\
  1.03  &   108   &  05 39 57.370 & -02 10 41.98 &    15.545 &      0.007 &  OO &    1.675 &      0.011 &  OO &    3.59 &  0.18 & 29.15 & 1.93 & 100.0 \\
  8.04  &   188   &  05 40 42.887 & -02 23 47.43 &    15.577 &      0.005 &  OO &    1.413 &      0.007 &  OO &    3.39 &  0.35 & -32.22 & 2.00 & 0.00\\
  3.01  &    51   &  05 38 23.070 & -02 36 49.32 &    15.716 &      0.014 &  OO &    1.423 &      0.022 &  OO &    1.57 &  0.27 & 34.50 & 2.51 & 92.1 \\
  2.03  &    63   &  05 40 34.389 & -02 44  9.52 &    15.770 &      0.007 &  OO &    1.850 &      0.011 &  OO &    3.28 &  0.17 & 33.15 & 1.24 & 100.0 \\
  4.03  &    29   &  05 38 22.824 & -02 45 30.43 &    15.876 &      0.006 &  OO &    1.314 &      0.010 &  OO &    3.34 &  0.35 & -10.71 & 7.19 & 0.00 \\
  4.03  &   368   &  05 38 26.833 & -02 38 46.04 &    16.123 &      0.006 &  OO &    1.911 &      0.011 &  OO &    2.22 &  0.30 & 30.38 & 3.21 & 99.6 \\
  2.03  &   191   &  05 40 29.437 & -02 40 55.91 &    16.229 &      0.005 &  OO &    1.368 &      0.007 &  OO &    3.48 &  0.25 & XC failed & - & - \\
  8.01  &   333   &  05 40 52.870 & -02 38 23.49 &    16.246 &      0.005 &  OO &    1.558 &      0.008 &  OO &    2.46 &  0.47 & 29.98 & 3.38 & 99.4 \\
  4.04  &   481   &  05 38 36.374 & -02 47  8.22 &    16.260 &      0.006 &  OO &    1.527 &      0.011 &  OO &    5.35 &  0.45 & 11.41 & 8.52 & 2.0  \\
  1.02  &   237   &  05 38 58.168 & -02 21 11.70 &    16.283 &      0.007 &  VV &    1.443 &      0.012 &  VV &    2.32 &  0.24 & -10.48 & 2.10 & 0.00 \\
  4.03  &   215   &  05 38 38.589 & -02 41 55.86 &    16.472 &      0.006 &  OO &    1.804 &      0.011 &  OO &    2.65 &  0.39 & 31.11 & 4.10 & 97.3 \\
  1.01  &   343   &  05 40 23.389 & -02 28 27.51 &    16.527 &      0.005 &  OO &    1.551 &      0.008 &  OO &    5.02 &  0.41 & 2.29 & 3.06 & 0.00 \\
  1.01  &   348   &  05 39 36.324 & -02 28  8.18 &    16.808 &      0.008 &  OO &    1.363 &      0.013 &  OO &    3.37 &  0.84 & -9.63 & 17.60 & 0.4 \\
  1.02  &    87   &  05 39  8.076 & -02 31 32.22 &    16.811 &      0.008 &  OO &    1.567 &      0.013 &  OO &    3.05 &  0.24 & -42.90 & 5.12 & 0.0 \\
  1.02  &   157   &  05 39 25.246 & -02 27 48.15 &    16.856 &      0.008 &  OO &    1.550 &      0.013 &  OO &    3.64 &  0.28 & 28.86 & 2.47 & 99.8 \\
  3.01  &   480   &  05 38 38.888 & -02 28  1.63 &    16.989 &      0.008 &  OO &    1.777 &      0.013 &  OO &    3.42 &  0.43 & 35.94 & 5.78 & 60.1 \\
  8.04  &    77   &  05 41 25.629 & -02 30 38.53 &    17.050 &      0.007 &  OO &    2.028 &      0.012 &  OO &    6.83 &  0.85 & - & - & - \\
  1.03  &   110   &  05 39 56.919 & -02 10 38.35 &    17.102 &      0.009 &  OO &    1.674 &      0.015 &  OO &    6.13 &  1.29 & 18.32 & 9.70 & 20.9 \\
  8.04  &   185   &  05 41 45.472 & -02 24 16.24 &    17.102 &      0.007 &  OO &    1.937 &      0.012 &  OO &    0.56 &  1.47 & - & - & - \\
  1.01  &   237   &  05 39 36.726 & -02 31 58.88 &    17.249 &      0.009 &  OO &    1.485 &      0.015 &  OO &    5.11 &  0.65 & -9.78 & 3.46 & 0.0 \\
  8.04  &   128   &  05 41 12.228 & -02 27 34.40 &    17.285 &      0.007 &  OO &    1.787 &      0.013 &  OO &    1.54 &  1.25 & - & - & - \\
  2.03  &   233   &  05 39 40.571 & -02 39 12.32 &    17.293 &      0.009 &  OO &    1.597 &      0.015 &  OO &    2.80 &  0.67 & 39.84 & 6.23 & 25.9 \\
  4.03  &   285   &  05 38 44.487 & -02 40 37.65 &    17.297 &      0.008 &  OO &    2.127 &      0.017 &  OO &    1.90 &  0.99 & - & - & - \\
  1.01  &   581   &  05 39 30.077 & -02 33 16.14 &    17.337 &      0.009 &  OO &    1.516 &      0.015 &  OO &    4.25 &  0.85 & -18.90 & 3.52 & 0.00 \\
  2.03  &   336   &  05 40 18.589 & -02 36 14.57 &    17.392 &      0.006 &  OO &    1.519 &      0.009 &  OO &    4.69 &  0.70 & -12.86 & 1.91 & 0.00 \\
  8.02  &   412   &  05 39 49.737 & -02 23 51.80 &    17.484 &      0.007 &  OO &    1.716 &      0.011 &  OO &    6.49 &  0.50 & 15.06 & 5.34 & 1.0 \\
  8.04  &    50   &  05 41 10.419 & -02 31 34.50 &    17.541 &      0.008 &  OO &    1.908 &      0.019 &  OO &    4.32 &  1.31 & - & - & - \\
  1.03  &   460   &  05 40 48.927 & -02 08 12.06 &    17.555 &      0.013 &  OO &    2.517 &      0.033 &  OO &    1.70 &  0.65 & - & - & - \\
  1.03  &   557   &  05 40  7.241 & -02 04  4.41 &    17.603 &      0.011 &  OO &    1.723 &      0.019 &  OO &    2.76 &  0.60 & -23.74 & 6.44 & 0.00 \\
  1.03  &   493   &  05 40  1.742 & -02 06 28.83 &    17.814 &      0.012 &  OO &    1.620 &      0.020 &  OO &    3.99 &  0.74 & 13.57 & 15.79 & 16.2 \\
  1.02  &   366   &  05 39  2.030 & -02 35 30.30 &    18.062 &      0.023 &  OV &    1.925 &      0.077 &  OV &    4.52 &  1.13 & -0.73 & 7.41 & 0.00 \\
  8.04  &   225   &  05 41  4.584 & -02 32 25.98 &    18.079 &      0.010 &  OO &    2.069 &      0.023 &  OO &    2.55 &  0.68 & 27.42 & 2.07 & 99.4 \\
  1.03  &   342   &  05 40  3.305 & -02 12 20.57 &    18.114 &      0.014 &  OO &    1.750 &      0.026 &  OO &    3.70 &  0.38 & 1.61 & 2.51 & 0.00 \\
  8.03  &   147   &  05 41  1.913 & -02 19  3.99 &    18.203 &      0.011 &  OO &    1.958 &      0.024 &  OO &    1.11 &  0.88 & - & - & - \\
  2.03  &   617   &  05 40 24.775 & -02 38 10.88 &    18.336 &      0.009 &  OO &    1.989 &      0.019 &  OO &    1.39 &  0.53 & 32.26 & 6.36 & 82.4 \\
  1.03  &  1094   &  05 40 36.238 & -02 04 49.07 &    18.452 &      0.022 &  OO &    2.052 &      0.047 &  OO &    1.70 &  0.96 & - & - & - \\
  1.02  &   672   &  05 39  7.958 & -02 15 43.04 &    18.540 &      0.019 &  OO &    1.709 &      0.036 &  OO &    1.00 &  0.59 & - & - & - \\
  1.01  &   800   &  05 40 12.537 & -02 26 16.49 &    18.821 &      0.013 &  OO &    1.838 &      0.027 &  OO &    1.31 &  1.35 & - & - & - \\
  1.03  &  1029   &  05 39 51.156 & -02 06  5.43 &    18.919 &      0.027 &  OO &    1.894 &      0.056 &  OO &    1.06 &  0.63 & - & - & - \\
  1.04  &  1100   &  05 40 19.740 & -02 16 54.11 &    18.976 &      0.027 &  OO &    2.182 &      0.066 &  OO &    7.93 &  1.00 & 31.09 & 11.66 & 57.5 \\
  1.03  &   612   &  05 40 33.842 & -02 13 42.97 &    19.295 &      0.038 &  OO &    2.097 &      0.088 &  OO &    2.52 &  0.77 & - & - & - \\
  8.03  &   457   &  05 40 58.366 & -02 11 34.65 &    19.333 &      0.029 &  OO &    2.407 &      0.079 &  OO &    -1.82 &  0.69 & - & - & - \\
  8.03  &   396   &  05 41  4.416 & -02 14 47.07 &    19.411 &      0.028 &  OO &    2.146 &      0.070 &  OO &    2.03 &  0.88 & - & - & - \\
  1.03  &   933   &  05 40 30.717 & -02 08  6.78 &    19.573 &      0.053 &  OO &    2.263 &      0.132 &  OO &    2.22 &  1.19 & - & - & - \\
  8.02  &  2051   &  05 39 54.188 & -02 26 25.25 &    19.709 &      0.031 &  OO &    2.010 &      0.070 &  OO &    -1.93 &  1.17 & - & - & -\\

\end{tabular}
\end{table*}

\subsection{Membership Probabilities}
\label{sec:prob}
\begin{figure} 
\includegraphics[height=300pt,width=225pt, angle=90]{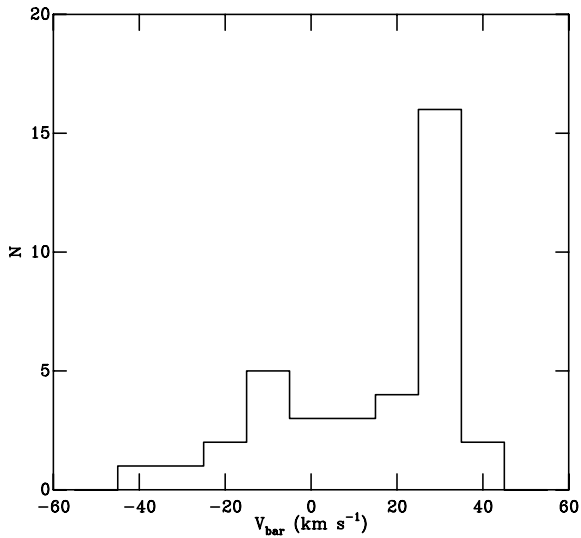}
\caption{A histogram of the velocities measured from our sample.}
\label{fig:vhist}
\end{figure}

\begin{figure} 
\includegraphics[height=300pt,width=225pt, angle=90]{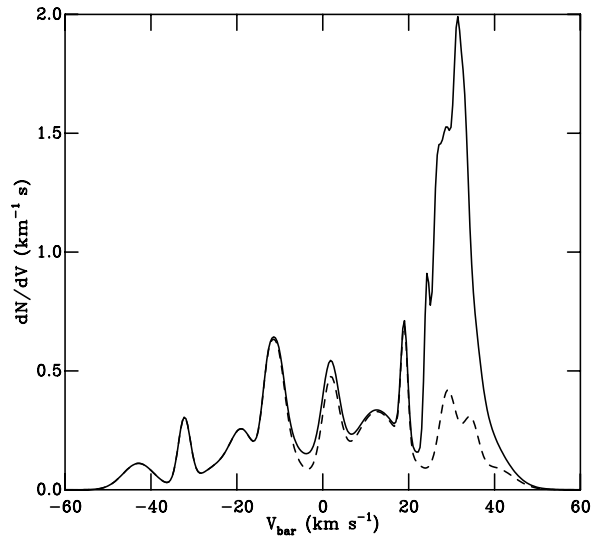}
\caption{A plot of the cumulative probability distribution of our
  sample.  The strongest peak is centered at 29.5 km s$^{-1}$.  The
  dotted line traces the cumulative probability distribution of the
  objects in the ``background'' region of the CMD (see Figure~\ref{fig:cmd}).}
\label{fig:cumprob}
\end{figure}

A histogram of the velocities obtained in the previous Section
(Figure~\ref{fig:vhist}) shows that there is no simple velocity cut
that will separate members of the $\sigma$ Ori group from
non-members.  Also the velocities derived in the previous Section have
a range of uncertainties, due the range of signal-to-noise in the different
spectra, so a simple  n$\sigma$-cut from a mean cluster velocity will
tend to preferentially select poor signal-to-noise objects as members.
Instead we can assign a probability of membership,
P$_{mem}$, to each object for which we have a velocity.   
This probability is the product of the probability that the object is
at an appropriate velocity to be considered a member, P$_{vel}$, and the
probability that an object at that velocity is not a radial velocity
contaminant, (1 - P$_{cont}$).  
The first question that must be answered, then, is what
constitutes an appropriate velocity for membership of the cluster?  
We have plotted the cumulative probability distribution of
the sample by summing Gaussians constructed from the velocities and
uncertainties for all objects with a radial velocity.   
As can be seen in Figure~\ref{fig:cumprob}, there is a
strong peak in this distribution centered at 29.5 km s$^{-1}$.  Since
we do not resolve the velocity dispersion of the cluster, we can take
the full width at the level of the background as the range of
velocities occupied by cluster members.  P$_{vel}$ is then simply
found as the fraction of the Gaussian derived from an object's
velocity and error that lies within the cluster range.  We take the
cluster range to be 24-37 km s$^{-1}$, based on the peak's full width at
the level of the background.  

The value we find for the group
velocity, 29.5 km s$^{-1}$ (and the range of 24-37 km s$^{-1}$), 
is consistent with that measured by   \citet{kenyon2004} of
31.2 km s$^{-1}$, and with the mean of values   for members of this
group measured by \citet{mhcbh2003} of 30.9   km s$^{-1}$.  
It is also consistent with the mean radial velocity of
  $\sigma$ Orionis itself of 29.2 $\pm$2.0 km s$^{-1}$
  \citep{wilson53}.  This is at odds with the value measured by
  \citet{zo2} of 37.3 km s$^{-1}$, but it is not clear why they have
  measured such a high velocity for this young group.
 
The probability that an object at the velocity of the cluster is a
radial velocity contaminant can be taken as the ratio of the
probability in Figure~\ref{fig:cumprob} integrated over a velocity
interval equal in size to the cluster range (but lying in the
non-cluster region) to the probability integrated over 
the cluster peak in Figure~\ref{fig:cumprob}. We find P$_{cont}$ = 0.2
for this sample.
We emphasise that this probability applies to the entire sample,
i.e. over the entire C-M space examined.  The value of P$_{cont}$ 
will change if, for example, one limits ones attention to the ``PMS''
region of the CMD (see below).  As such we only list values of P$_{vel}$ in
table~\ref{table:results} so as to avoid misinterpretation of the results.

\begin{figure} 
\includegraphics[height=375pt,width=275pt, angle=90]{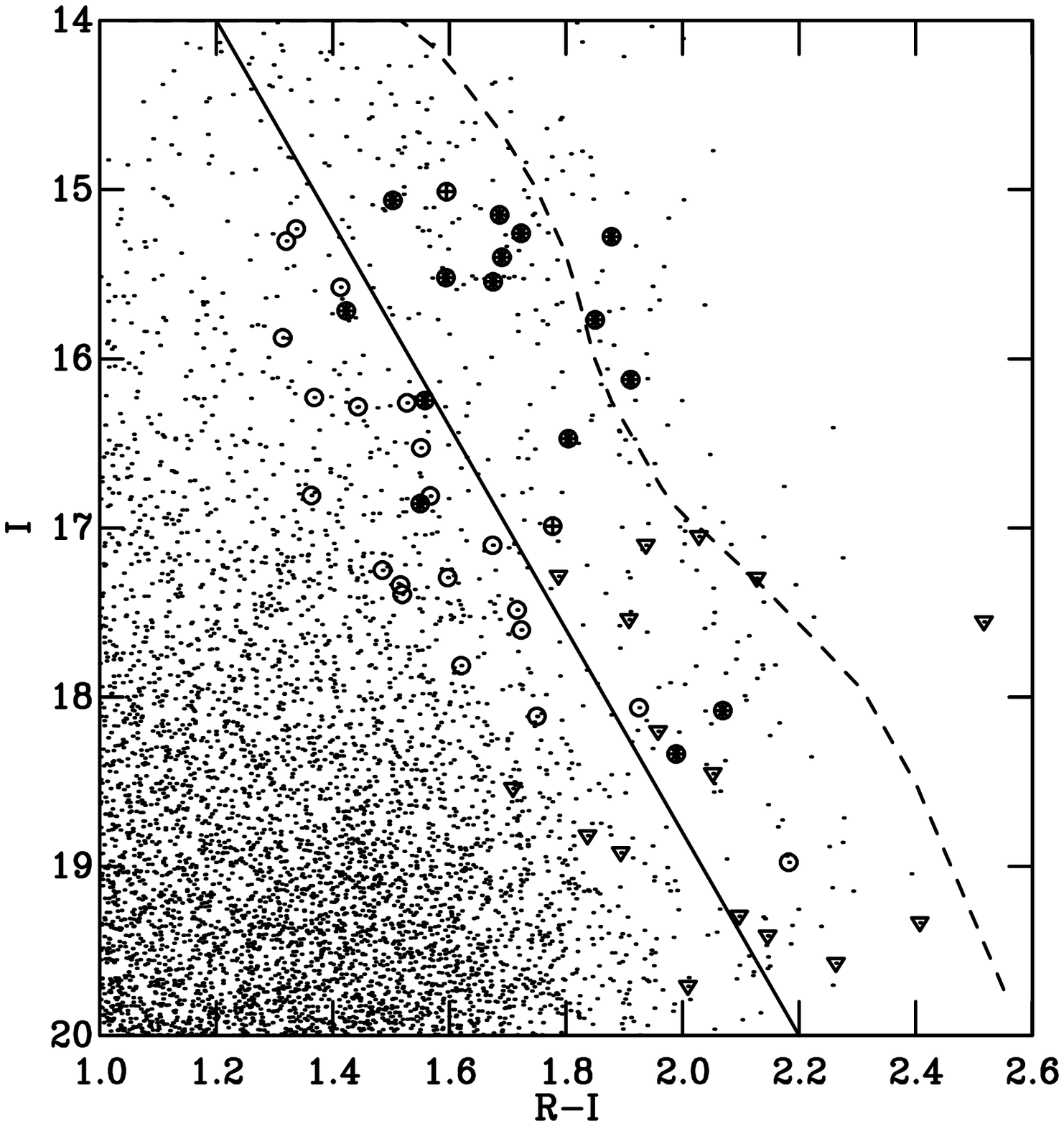}
\caption{The CMD as in Figure~\ref{fig:sample}.  Objects with detected
 Na{\sc i} are shown as open circles.
Objects with P$_{vel} >$ 80\% are filled circles,
  whilst circles filled with crosses are objects with P$_{vel} >$60\%.  
Objects for which no Na{\sc i} was detected
 are shown with open triangles.  The dotted line follows a NextGen 5
 Myr isochrone \citep{cb97, bcah2002}. Our expected PMS region is
 defined as redward of the solid line.}
\label{fig:cmd}

\end{figure}

\section{Discussion}
\label{sec:disc}

In the following discussion we refer to the region redward of the
solid line in Figure~\ref{fig:cmd} as the expected PMS region.  The
region that lies blueward of this line is referred to as the
background region.  This locus was selected as the dividing line as it
corresponds to a liberal photometric selection, and so provides an
interesting case for investigating exclusion of members and
contamination by non-members.

\subsection{Are there members outside the ``PMS'' region?}

The simplest way to answer this question is to examine the cumulative
probability distribution of the velocities for the objects in the
``background'' region of the CMD.  We have plotted this distribution with a
dotted line in Figure~\ref{fig:cumprob}.  It is clear from this plot
that there is no over-density in the probability distribution at the
cluster velocity.  This implies that there is not a significant number of
members in the background region of the CMD, and hence photometric
selection techniques are not excluding members
of the $\sigma$ Ori young group.  This is the principal conclusion of
this work.

\subsection{Contamination within the ``PMS'' region}

As can be seen in Figure~\ref{fig:cmd}, at magnitudes brighter than
about I = 17 we find that all of the objects in the expected PMS region have
P$_{vel} >$ 80\% except one, which has P$_{vel} =$ 77.3\%.   
This is consistent with the result of
\citet{kenyon2004} that essentially all objects in the
PMS region of the CMD are members of the $\sigma$ Ori young group.

Fainter than I = 17 the situation is somewhat different. We
only detect Na{\sc i} in 13 out of 29 objects in this part of our sample.  
The failure to detect Na{\sc i} in so many objects in this part of the CMD
makes a discussion of the contamination problems in this
region impossible.  However, this failure is itself worthy of some
discussion.  As already mentioned, the signal-to-noise threshold used
to select our sample was chosen by comparison to \citet{mdg2004}, who
find EW(Na{\sc i}) $\geq$ 3~\AA\ for all but 2 of their members of the
5 Myr old Upper Scorpius association. As such we expected that the
majority of members of the similar aged $\sigma$ Ori
group might share this property. 
By comparing the EWs of members in both associations we can assess any
differences between them. 
It would be simplest to compare the EWs of likely members in the two
associations using $J-I$ colours, since this is a fair proxy to
spectral type, however 2MASS colours are only available for the 10
brightest of our likely members.
Instead we compare the EWs in the two samples using the I-band
magnitudes, corrected for the differing distances to the two
associations.

\begin{figure} 
\includegraphics[height=230pt,width=175pt, angle=90]{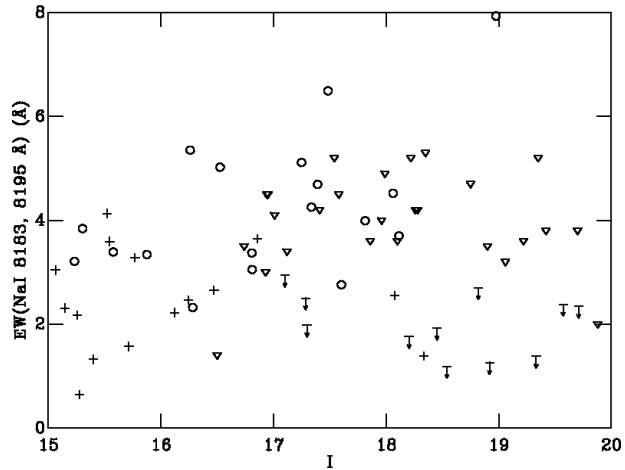}
\caption{A plot of EW(Na{\sc i}) versus I-band magnitude for our objects
  with $P_{vel} > 80\%$ (crosses), $P_{vel} < 60\%$ (open circles) and 
\citet{mdg2004} members of Upper  Scorpius (triangles).  
The magnitudes for the Upper Scorpius objects
  have been scaled for the distance to $\sigma$ Ori.  Also plotted are
  EW(Na{\sc i}) upper limits for the 11 objects for which our
  objective measure failed to detect Na{\sc i}.}
\label{fig:ew}
\end{figure}

Figure~\ref{fig:ew} shows that the likely members from our sample display
weaker Na{\sc i} EWs than the \citet{mdg2004} members of Upper Scorpius.
This is not surprising in the case of the objects brighter than I = 17
as they will be of earlier spectral type than any of the \citet{mdg2004}
objects, and the Na{\sc i} doublet gets weaker with increasing temperature
at a given value of $log(g)$ \citep{sbrm97}.  By this token, we might
expect the fainter objects, presumably with later spectral types, to
display stronger Na{\sc i} absorption, however this does not seem
to be the case.  Both of the likely members with I $>$ 17 have
EW(Na{\sc i}) $<$ 2.5~\AA.
Inspection of the measured EWs for the failed
detections (see Table~\ref{table:results}) indicates that the majority
of them are consistent with values of EW(Na{\sc i})$<$ 2~\AA.  This is
somewhat lower than the typical values seen in the \citet{mdg2004} sample.
Thus we are presented with two options: 1) the failed detections are
interloping giants and sub-giants with low surface gravities; 2) many
of the failed detections are member brown dwarfs which display weak
Na{\sc i} absorption. 
Option 1 requires many giant stars to be located at great distance and
along our line of sight, yet
with colours that coincide with the expected PMS region of
C-M space. Since the line of sight towards $\sigma$ Ori is
elevated -17.3$^\circ$ from the galactic plane, it is hard to see that
option 1 is feasible. 
Option 2, however, requires
the low-mass objects in $\sigma$ Ori to display consistently
weaker Na{\sc i}
absorption in the 8183, 8195~\AA\ doublet than has been observed in Upper
Scorpius by \citet{mdg2004}.

Upper Scorpius is thought to be marginally older than $\sigma$
Ori, 5-10 Myrs versus 1-5 Myrs, so we might expect to see weaker Na{\sc i}
absorption in $\sigma$ Ori, as $log(g)$ will be lower in the younger
group.
\citet{mbjaha2004} derive surface gravities for a number of low-mass
objects in Upper Scorpius, and compare them to model predictions.
  They find that $log(g)$ changes from about 3.7 to 4.2 between 1 and 10 Myrs
  for a 0.04 M$_\odot$ 
  brown dwarf (I = 18 at the distance of $\sigma$ Ori), and even less
  of a change at lower masses.   
Given that giants have $log(g) \approx$ 0, and display EW(Na{\sc i})
$\approx$ 1~\AA \citep{sbrm97},  it is difficult to see how such a
small change in $log(g)$ could give rise to the factor 2 discrepancy
between the EW(Na{\sc i}) measured here for objects in $\sigma$ Ori and those
measured for objects in Upper Scorpius by \citet{mdg2004}.  
The discrepancy between the two sets of EWs can likely be explained
without resorting to a physical interpretation.  The spectra used by
\citet{mdg2004} are significantly lower resolution (R=900) than those
used here, and this could lead to placing the pseudo-continuum over
blended absorption features such as the TiO band just redward of the
Na{\sc i} doublet.
Thus we suggest that option 2 is the most likely
explanation for these results.
 We can not determine if this is due to a physical difference between
 the two regions or if it has resulted from a difference in
 measurement techniques.
  
This result implies that it will be very hard to
distinguish between background giants and member brown dwarfs based on
the strength of Na{\sc i} absorption.

The low values of EW(Na{\sc i}) implied for faint objects in our sample
suggest that to  avoid biasing such samples against finding members,
observations should be made to a sufficient signal-to-noise to detect
the Na{\sc i} doublet in giant stars. 
Unfortunately, this destroys one of the main
benefits of using the Na{\sc i} doublet for radial velocity determination
in main sequence stars, the ability to measure it at low
signal-to-noise due to its strength.  This is offset, however, by the
fact the gravity sensitive nature of this doublet
can provide a second diagnostic for contaminants

A possible weakness in use of radial velocities to define membership
of young groups is the fact that bona fide members can be ruled out
when they are members of binary systems.  \citet{dm2001} use the 
radial velocities in concert with the presence of Lithium absorption
at 6708~\AA\ to identify 266 likely members of the $\lambda$ Ori
young group.  They find that 9 objects with EW(Li) consistent with
youth are double lined binaries, while 3 appeared to be single line
binaries.  Whilst double lined binaries can be recognised as such, and
not ruled out of membership lists, the single lined binaries would be
ruled out in the absence of additional evidence of their membership.  
If we assume a similar binary fraction
and distribution of properties as that seen by \citet{dm2001} in
$\lambda$ Ori, we can
estimate that for our sample of 18 likely members, we would not expect
to identify any double lined binaries, or to miss any single lined
ones.  
Since we cannot use EW(Na{\sc i}) to identify members, we cannot
assess whether we have ruled out any single lined binaries from our
membership lists. However, a
significant departure from behaviour elsewhere in the Orion OB1
association would be required for a
significant number of members to be ruled out due to binarity.

We have cross-correlated our sample against the candidate membership lists of
\citet{bejar99,bejar2001}.  We find that only one object,
3.01 480, correlates in position with an object, SORI 68, from their
sample.  
However, we do not believe that they are the same object since there
is a nearly 6 magnitude discrepancy in their I-band brightness
(I=16.99 vs. I=23.78).  The fact that we do not share objects with the
\citet{bejar99,bejar2001} sample is explained by a number of factors.
Firstly, our $RI$ catalogue excludes the regions immediately adjacent to
the $\sigma$ Ori multiple system, whereas the B\'ejar et al sample does
not. In addition, our catalogue covers a much larger Section of sky
\citep{kenyon2004}, so there are many objects that B\'ejar et al were
simply unable to select as candidates.
Finally, we draw our sample from a wider region of C-M space
than B\'ejar et al, and so we simply miss some of their candidates due
to a lower density of selected objects.

\section{Conclusions}
\label{sec:conc}

We have carried out high resolution spectroscopy of the Na{\sc i} doublet
at 8183, 8195~\AA\ of a sample of candidate low-mass stellar and substellar
 members of the $\sigma$ Ori young group drawn from a broad region
 of C-M space.  We have selected a sample of 54 objects
 which displayed sufficient signal-to-noise to detect Na{\sc i} with
 EW(Na{\sc i}) = 3~\AA\ at a significance of 2$\sigma$, a criterion based on
 observations of brown dwarfs in Upper Scorpius by \citet{mdg2004}.  
Significant  (EW(Na{\sc i}) $>$ 2$\sigma_{EW}$) Na{\sc i} was detected in 38 of
the 54 sample  objects, and these were cross-correlated against an M9V
standard to obtain radial velocities, which were then used to
calculate membership probabilities.  We find that 13 objects are
likely radial velocity members (P$_{vel} \geq 80\%$) of the $\sigma$
Ori young group.    
Based on these probabilities, and the values measured for EW(Na{\sc i}) we
arrive at  the following conclusions.\\
\\
1) Photometric selection techniques do not miss significant
numbers of bona fide association members.\\
\\
2) At I brighter than 17 the expected PMS region of the CMD does not contain a
significant number of contaminants.\\
\\
3) Very low-mass objects in the $\sigma$ Ori young group appear to have
weaker EW(Na{\sc i}) than found by \citet{mdg2004} for low-mass members of
the Upper Scorpius OB association.  We have no explanation for why
this is so, although we suggest measurement effects could account for this.\\
\\
4) High resolution observations of the Na{\sc i} doublet at 8183, 8195~\AA\
offer the possibility of 2 membership diagnostics for very low-mass
objects from a single observation. 
Ensuring observations are made to sufficient
signal-to-noise to detect the doublet in the spectrum of a ($log g
\approx 0$) giant star will  avoid biasing a sample against bona fide
members with lower than expected EW(Na{\sc i}).\\ 
\\

\section*{Acknowledgements}
SPL is supported by PPARC.  The authors acknowledge the data analysis
facilities at Exeter provided by the Starlink Project which is run by
CCLRC on behalf of PPARC.  A huge `thank you' goes to Dr Alasdair Allan
for his work on fibsplitter and the web service.

\bibliography{refs}
\bibliographystyle{mn2e}

\end{document}

%% file: aas_macros.tex
%
%
%
%


\def\aj{\rm{AJ}}                   
\def\araa{\rm{ARA\&A}}             
\def\apj{\rm{ApJ}}                 
\def\apjl{\rm{ApJ}}                
\def\apjs{\rm{ApJS}}               
\def\ao{\rm{Appl.~Opt.}}           
\def\apss{\rm{Ap\&SS}}             
\def\aap{\rm{A\&A}}                
\def\aapr{\rm{A\&A~Rev.}}          
\def\aaps{\rm{A\&AS}}              
\def\azh{\rm{AZh}}                 
\def\baas{\rm{BAAS}}               
\def\jrasc{\rm{JRASC}}             
\def\memras{\rm{MmRAS}}            
\def\mnras{\rm{MNRAS}}             
\def\pra{\rm{Phys.~Rev.~A}}        
\def\prb{\rm{Phys.~Rev.~B}}        
\def\prc{\rm{Phys.~Rev.~C}}        
\def\prd{\rm{Phys.~Rev.~D}}        
\def\pre{\rm{Phys.~Rev.~E}}        
\def\prl{\rm{Phys.~Rev.~Lett.}}    
\def\pasp{\rm{PASP}}               
\def\pasj{\rm{PASJ}}               
\def\qjras{\rm{QJRAS}}             
\def\skytel{\rm{S\&T}}             
\def\solphys{\rm{Sol.~Phys.}}      
\def\sovast{\rm{Soviet~Ast.}}      
\def\ssr{\rm{Space~Sci.~Rev.}}     
\def\zap{\rm{ZAp}}                 
\def\nat{\rm{Nature}}              
\def\iaucirc{\rm{IAU~Circ.}}       
\def\aplett{\rm{Astrophys.~Lett.}} 
\def\apspr{\rm{Astrophys.~Space~Phys.~Res.}}
\def\bain{\rm{Bull.~Astron.~Inst.~Netherlands}} 
\def\fcp{\rm{Fund.~Cosmic~Phys.}}  
\def\gca{\rm{Geochim.~Cosmochim.~Acta}}   
\def\grl{\rm{Geophys.~Res.~Lett.}} 
\def\jcp{\rm{J.~Chem.~Phys.}}      
\def\jgr{\rm{J.~Geophys.~Res.}}    
\def\jqsrt{\rm{J.~Quant.~Spec.~Radiat.~Transf.}}
\def\memsai{\rm{Mem.~Soc.~Astron.~Italiana}}
\def\nphysa{\rm{Nucl.~Phys.~A}}   
\def\physrep{\rm{Phys.~Rep.}}   
\def\physscr{\rm{Phys.~Scr}}   
\def\planss{\rm{Planet.~Space~Sci.}}   
\def\procspie{\rm{Proc.~SPIE}}   

\let\astap=\aap
\let\apjlett=\apjl
\let\apjsupp=\apjs
\let\applopt=\ao